\documentclass{eptcs}
\providecommand{\event}{SOS 2007} 
\usepackage{graphicx}
\title{Towards a General Framework for Formal Reasoning about Java Bytecode Transformation}
\author{Razika Lounas
\institute{University of Boumerdes,Algeria}
\email{razika\_lounas@umbb.dz} \and Mohamed Mezghiche
\institute{University of Boumerdes,Algeria}
\email{mohamed.mezghiche@umbb.dz} \and Jean-Louis Lanet
\institute{University of Limoges, France}
\email{jean-louis.lanet@xlim.fr} }
\def\titlerunning{Towards a General Framework for Formal Reasoning about Java Bytecode Transformation}
\def\authorrunning{Razika Lounas, Mohamed Mezghiche \& Jean-Louis Lanet}
\begin{document}
\maketitle

\begin{abstract}
Program transformation has gained a wide interest since it is used
for several purposes: altering semantics of a program, adding
features to a program or performing optimizations. In this paper
we focus on program transformations at the bytecode level. Because
these transformations may introduce errors, our goal is to provide
a formal way to verify the update and establish its correctness.
The formal framework presented includes a definition of a formal
semantics of updates which is the base of a static verification
and a scheme  based on Hoare triples and weakest precondition
calculus to reason about behavioral aspects in bytecode
transformation.
\end{abstract}

\section{Introduction}

Program transformation is a technique used for several proposes:
altering semantics of a program, performing optimizations or
adding features. Several tools were developed in this domain, for
example, Java Syntactic Extender (JSE)
\cite{DBLP:conf/oopsla/BachrachP01} and BCEL
\cite{DBLP:conf/jit/Dahm99} for Java. However, in some cases, the
source code is not available (or not distributed). Transforming a
program at bytecode level is an interesting alternative since
several languages like Java, Java Card or $C\sharp$ are based on
virtual machines executing bytecode. Besides, in transformations
at bytecode level we don't need recompiling (which may take time)
as in the case of transformations at source code level. On the
other hand, bytecode level transformation is more complex than
source-level manipulation to the users because they have to know
bytecode language very well and because of the many low-level
details one needs to deal with, in contrast with source code
level.

Bytecode transformation is used in several applications. In
\cite{DBLP:conf/ma/SakamotoSY00}, the authors developed an
algorithm to ensure portable thread migration in Java. This
algorithm is based on bytecode transformation. Bytecode is
transformed in order to enable programs to save and restore their
execution state after migration through the network. Another
purpose for bytecode transformation is presented in
\cite{Binder:2005:JBT:1705549.1706046} where a framework based on
bytecode transformation is developed in order to enable Java
applications to perform CPU management

Transforming a program may occur at runtime. The update is then
said to be dynamic (Dynamic Software Update: DSU). In
\cite{phdthesis1} \cite{DBLP:conf/icde/NoubissiIL11}, the authors
presented a system to perform dynamic software update: while the
Java Card virtual machine is executing the program, the bytecode
is updated. In \cite{Denker:2006:RBT:1646593.1646614} , a tool is
developed in order to perform runtime bytecode update for
Smalltalk.

This large interest of bytecode transformation and its use in many
applications lead to the question of its correctness. In fact, a
transformation may introduce an error which may alter the bytecode
in a different way from that is expected by the programmer. In
addition, some applications where the update occurs are critical,
such as in Java Card. In these applications where security issues
are involved the update must pass certification procedure for
example Common Criteria \cite{bibliographystylewebpage1} . For a
certain certification level one has to provide a formal proof of
the security mechanism implemented. A formal way to reason about
transformations and verify their validity is then necessary.

In this work, we present a first step for a general framework for
reasoning about bytecode transformation. We focus on Java bytecode
and the system presented in \cite{DBLP:conf/icde/NoubissiIL11}
called embedDSU: a system to update dynamically Java Card
applications. But this is not restrictive: the framework developed
may be applied to other systems and for this it is general. The
framework is divided in two parts: we propose an approach for a
static analysis by defining a formal semantics for update to
ensure the absence of type errors and then in the second part we
propose an approach to reason about behavioral aspects using Hoare
triples and predicate transformations.

This paper is organized as follow: in section 2 we give an
overview of embedDSU. Section 3 introduces a verification approach
through a static analysis of the bytecode. In section 4 we present
the part of the framework which talks about reasoning on the
behavioral aspects of updates. We present related work in section
5 and conclude in section 6.

\section{Overview of EmbedDSU}

EmbedDSU \cite{phdthesis1}
\cite{10.1109/ICONS.2010.27}\cite{DBLP:conf/icde/NoubissiIL11}, is
a software-based DSU technique for Java based smart cards which
relies on the Java virtual machine. It is based on the
modification of an embedded virtual machine. EmbedDSU is divided
in two parts: off-card and on-card:
\begin{itemize}
 \item In off-card, in
order to apply the update only to the parts of the application
that are really affected by the update, a module called DIFF
generator determines the syntactic changes between versions of
classes. The changes are expressed using a Domain Specific
Language (DSL). Then, the DIFF file result is transfered to the
card and used to perform the update.
 \item The on-card part is divided
into two layers: 1) Application Layer: The binary DIFF file is
uploaded into the card. After a signature check with the wrapper,
the binary DIFF is interpreted and the resulting instructions are
transferred to the Patcher in order to perform the update. The
Patcher has the role of initializing update data structures. These
data structures are read by the updater module to determine what
to update and how to update, by the safeUpdatePoint detector
module to determine when to apply the update and by the rollbacker
to determine how to return to the previous version in case of
update failure. All these issues pass through the introspection of
the virtual machine.
 2)System Layer: The modified virtual machine supports the followings
features: (1) Introspection module which provides search functions
to go through VM data structures like the references tables, the
threads table, the class table, the static object table, the heap
and stack frames for retrieving information necessary to other
modules; (2) updater module which can modify and update object
instances, method bodies, class metadata, references, affected
registers in the stack thread and affected VM data structures; (3)
SafeUpdatePoint detector module which permits to detect safe point
in which we can apply the update by preserving coherence of the
system.
\end{itemize}
\begin{figure}[!ht]
\begin{center}
\includegraphics[width= 11cm]{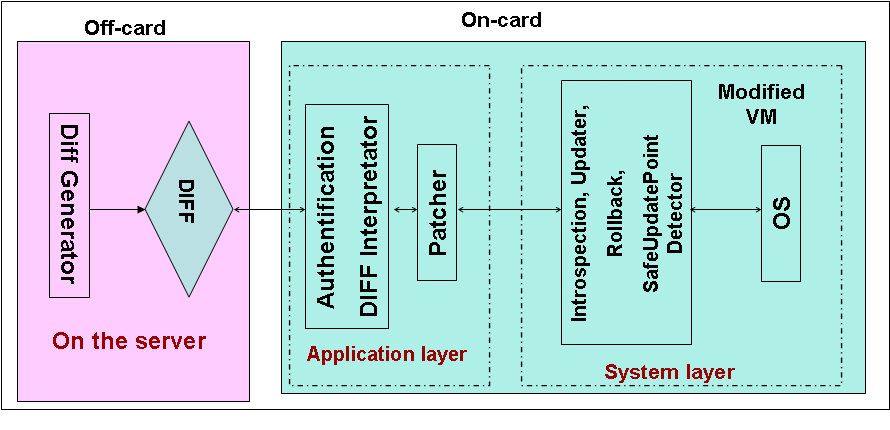}
\caption{Architecture of EmbedDSU}
\end{center}
\end{figure}
EmbedDSU updates three principal parts:
\begin{itemize}
\item The bytecode: The process updates first the bytecode of the updated class and the meta data associated with it:
constant pool, fields table, methods table...
\item The heap: The process updates the instances of the updated class in the heap, obtains
new references for modified objects and updates instances using
these references.
\item The frames: The process updates in each frame in the thread stack the references of updated objects to point to new instances.
\end{itemize}

This paper addresses the first part: bytecode update at the method
level. The types of updates that may occur are: adding, modifying
or suppressing bytecode instructions, changing the signatures of a
method or modifying local variables. These updates are contained
in the DIFF file (also called patch) which indicates exactly which
is the update and where it occurs in the bytecode. For example,
when adding an instruction, the patch informs the system which
instruction to add and where to add it (information about the
program pointer)

\section{Updated Bytecode analysis for static verification}

We present an approach for transformation validation based on
static semantics of bytecode (figure 2) in order to avoid type
errors in transformed programs. From a first version$£BC\_V1$ and
a second version $BC\_V2$ (Version one transformed), we have a
DIFF file. This DIFF file is applied to the first version. We
obtain a version $BC\_V1.2$ (annotated $BC\_V1$ on the figure).
The goal of the validation is to establish that $BC\_V1.2$ and
$BC\_V2$ are semantically equivalent by comparing $V\_Sem \  S1$
and $V\_Sem \  S2$ representing the semantics of $BC\_V1.2$ and
$BC\_V2$ respectively.

\begin{figure}[!ht]
\begin{center}
\includegraphics[width= 11cm]{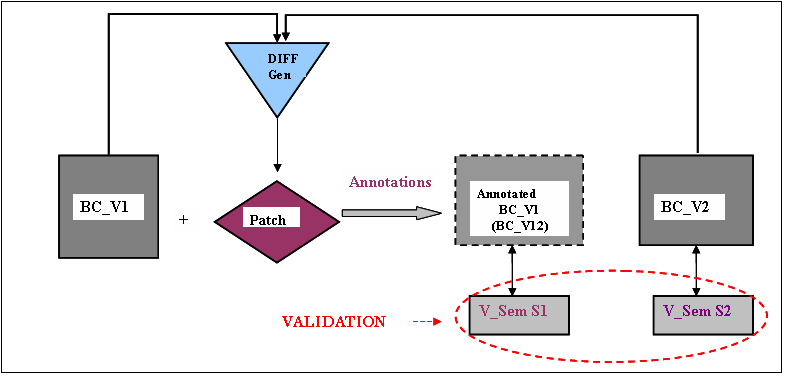}
\caption{Approach for static verification}
\end{center}
\end{figure}
The application of the DIFF to the first version is modeled
syntactically as annotations (figure 3).
 We insert annotations to indicate instructions addition and suppression. For example,
 \verb"Del \%2" : deletes the instruction at program counter $(pc)$ 2  and \verb"add \%6 inc", adds the instruction inc at $pc$  6.

 \subsection{The language}

For  the definition of the  static semantics, we adopt the
formalism used by Freund and Mitchell
\cite{Freund:1999:TSO:330643.330646}. The authors define a type
system for a small subset of Java bytecode. We extend this subset
with instructions to indicate updates called update instructions
(Upd\_instr)for instruction addition, deletion and modification.
In this definition, $x$ is a local variable; $L$ is an instruction
address; $A$ is a class name; $f$ is a field name; $l$ is a method
name and pc the program counter.
          \begin{verbatim}
Instruction::= | pop | if L| store x| load x| new A| goto L|inc
               |add |invokevirtual A l t|getfield  A f t|putfield A f t
Upd_Instr::= Add_Inst Instruction at pc
             |Dlt_Inst Instruction at pc |Mod_Inst Instruction at pc
                    \end{verbatim}

\begin{figure}[!ht]
\begin{center}
\includegraphics[width= 11cm]{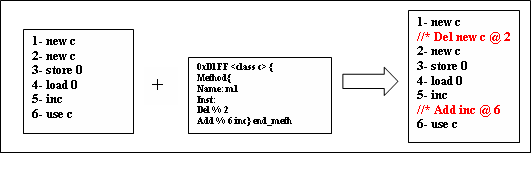}
\caption{Bytecode annotation with update instructions}
\end{center}
\end{figure}

\subsection{Formal semantics}

We propose a static semantics to express the effect of update
instructions on a configuration of the bytecode. In the four rules
shown in Fig 4 , $F$ is a mapping from a program point to a
mapping from a frame variable to a type. $S$ is a mapping from a
program point to an ordered sequence of types, $i$ denotes a
program point or an address of code. The map $F_{i}$ gives a type
of local variables at program point $i$. The string $S_{i}$ gives
the types of entries in the operand stack at program point $i$.
These $F$ and $S$ are useful to our semantics since they contain
typing information about valid local variables and entries in the
operand stack respectively. $SD$ represent the stack depth and $M$
(mapping) is a function that associate a number to each line.
$Dom$ is the set of addresses used by the method. A configuration
at line i is represented by $ <(F,S, SD, M),i >$.

\begin{figure}[!ht]
\begin{center}
\includegraphics[width= 13cm]{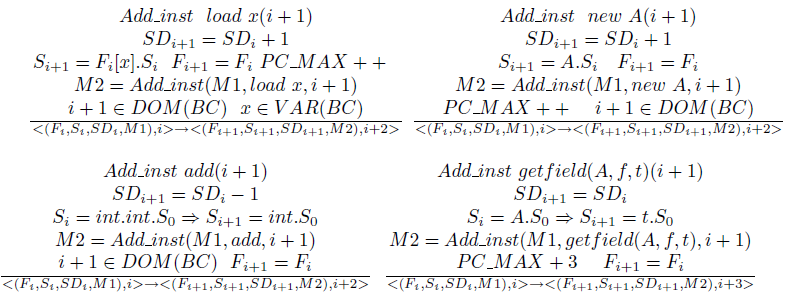}
\caption{Rules for update instructions}
\end{center}
\end{figure}

For illustration, the add of the instruction \verb"new A" at line
$i+1$ allows us to obtain a new configuration if the stack depth
is incremented, local variables are not affected and in the stack,
the type $A$ is inserted. $M2$ is the result of operations on
$M1$. These operations which represent manipulations on bytecode
are: range and shift. The operation range extracts from a mapping
$M1$ a part $M2$ included between line $n$ and line $m$. The
operation shift shifts a part from a mapping between $n$ and $m$
for $p$ positions. Both
operations are of type:\\
 $mapping* int* int*\rightarrow mapping$

In order to take into account jumps in bytecode transformation, we
define two other operations: look\_for\_jumps which returns from a
mapping a list of jumps instructions represented by their line
number and the operation update\_jumps to update jump
instruction:\\
$ Look\_for\_jumps:mapping \rightarrow int list\\
 Update\_jumps : mapping * int list* int \rightarrow mapping $

 Due to a matter of space, we don't give in this paper rules for
 ordinary bytecode instructions, rules for instruction suppression and the
 remaining rules for instruction addition.

 \section{An approach for reasoning about transformations}

In this section, we present an approach to reason about behavioral
aspects of transformations on bytecode. This approach is based on
bytecode specification in term of preconditions and postconditions
and on predicate transformation to generate verification
conditions. We give first some definitions before presenting the
scheme of the approach.

\subsection{Definitions}
\textbf{Definition 1. }\textbf{\emph{Hoare triplet}} A Hoare
triple is the basic object in Hoare logic
\cite{DBLP:journals/cacm/Hoare69} . It has the form of \{P\} S
\{Q\} where P and Q are logical formulas and S a program. The
interpretation of \{P\} S \{Q\} regarding partial correctness is:
If S is executed in a state in which P holds, then it terminates
in a state in which Q holds unless it aborts or runs forever. The
interpretation in total correctness is: if S is executed in a
state in which P holds, then it terminates in a state in which Q
holds.

 Reasoning in Hoare logic is based on inference rules. Here is an example of a general rule:
\begin{center}
 $
 \frac{P\Rightarrow P1 \ \ \{P1\} \ S \ \{Q1\} \ \ Q1\Rightarrow Q}
 {\{P\}\ S \ \{Q\}}
 $
\end{center}
\textbf{Definition 2. }\emph{\textbf{Weakest precondition (WP)
calculus}} The Weakest Precondition calculus
\cite{Dijkstra:1972:CIN:1243380.1243381}is a predicate transformer
that takes a
  code S and a postcondition Q and returns a precondition. We write WP(S, Q): "the weakest precondition of S regarding Q". WP(S,Q)
   is a precondition for S that ensures Q as a postcondition. It is weakest in the sense that if we take any P such that \{P\} S \{Q\}
    then $ P \Rightarrow WP(S, Q).$ It satisfies \{WP(S,Q)\} S
    \{Q\}.

\noindent \textbf{Definition3.}  \emph{\textbf{Strongest
postcondition (SP) calculus}} The Strongest Postcondition calculus
\cite{Dijkstra:1972:CIN:1243380.1243381}
 is a predicate
 transformer that takes a precondition P and a code S and returns a postcondition. We write SP(P, S) as
  "the strongest postcondition of S  regarding P". SP(P, S) is a postcondition for S that is ensured by precondition P. It
   is strongest in the sense that if we take any Q such that \{P\} S \{Q\} then $SP(P, S)\Rightarrow Q $. It satisfies \{P\} S \{SP(P, S)\}.

\subsection{Approach Description}

We propose an approach based on the definition of the concept of
\emph{\textbf{triple transformation}}. It represents the idea that
an update of an existing method M1 with precondition and
postcondition P1 and Q1 results of a new method M2 with a new
specification P2 and Q2. The triple
 \{P1\} M1 \{Q1\} is transformed via the update to a new triple \{P2\} M2 \{Q2\}. The approach defines these concepts: \emph{initial triplet},
  \emph{target triplet} and \emph{calculated triplet}:\\

\noindent \textbf{Definition 4.}\emph{\textbf{Initial triplet}} An
initial triple \{P1\} M1 \{Q1\} represents a method M1, its
precondition P1 and its postcondition Q1 at
 the initial state, that means before an update. This triple represents a method and its specification in the running
 code.\\

\noindent \textbf{Definition 5.} \emph{\textbf{Target triplet}} A
target triple \{P2\} M2 \{Q2\} represent a new version M2 of the
initial version M1 and its specifications P2 and Q2. It is the
goal of the update as it is written by the programmer. The methods
M1 and M2 are written in bytecode.
 Pre/post-conditions (P1, Q1, P2 and Q2) are written using existing specification languages and
 tools by the programmer.\\

\noindent \textbf{Definition 6}.\emph{\textbf{Calculated Triplet}}
A calculated triple is a triple obtained starting from an initial
triple with the application of the transformations contained in a
patch (list of update instruction). It is the result of the
transformation of an initial triple.It is calculated using the
Transform\_triple algorithm.

\begin{figure}[!ht]
\begin{center}
\includegraphics[width= 12cm]{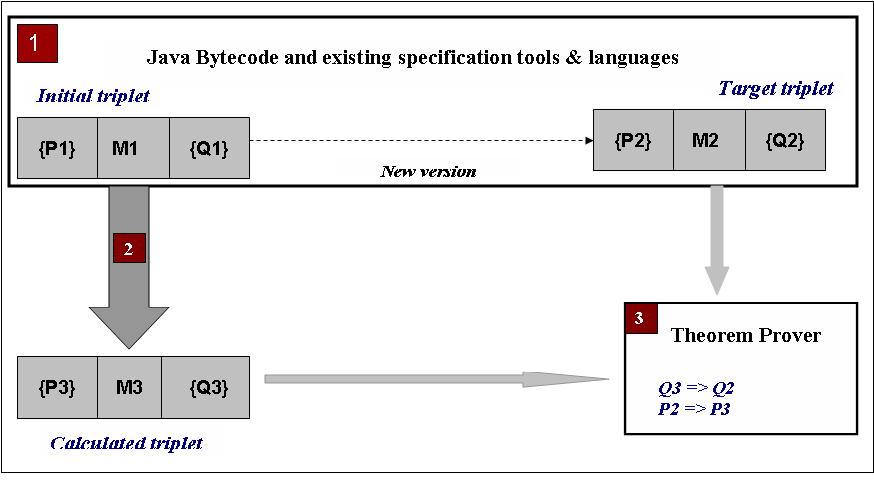}
\caption{The proposed Approach for reasoning}
\end{center}
\end{figure}

As shown in figure 5, the approach is based on three steps:
\begin{itemize}
\item \textbf{Step (1):} \emph{\textbf{Programming and
specification}} The initial code M1 is written in bytecode and the
specification (pre/post-condition)
 is written using existing specification language and tools.  The new version of M1 called M2 is written in bytecode. The desired specification
 of the update is expressed by the programmer using existing tools too and is expressed in term of pre/post-conditions of the new code M2.

\item \textbf{Step (2):} \emph{\textbf{Triple transformation}}
Given an initial triple and a list of update instructions
contained in a patch, this calculus transforms the initial triple
step by step. Each step correspond to the application of an update
instruction. We take the case of instruction insertion. The
application of an update instruction returns an intermediate
triple that will be taken as an argument of the calculus for the
next update instruction. This is represented as a recursive
algorithm called \emph{Transform\_triple}. It is based on the
predicate transformation calculus: weakest precondition (wp) and
strongest postcondition (sp).
 \begin{verbatim}
Transform_triple (p1, q1, m1, patch1) = match patch1 with
 |[]->return (p1,q1)
 | Add_instr (X,i)::patch2-> let n=last_line(m1) in
          let m2=m1(+)(X,i) in let wp1 = WP (m1[i,n], q1) in
           if wp1 != WP(m2[i+1,n], q1)
              then Raise Exception
           else
           let p2=WP(m2 [ 1,i ], wp1) in
           let sp1 = SP (m1 [1,i-1], p1) in
            if sp1!=SP (m2[1,i-1], p1)
             then Raise Exception
           else
            let q2 = SP (m2[I,n], sp1) in
             Transform_triple(p2,q2,m2,patch2)
 \end{verbatim}

The algorithm Transform\_triple represents the application of a
patch\emph{ patch1} to a method \emph{m1}with specification
\emph{p1} and \emph{q1}.The patch contains update instructions
about inserting instructions \emph{(X)}at an indicated line
\emph{i} (ADD\_instr (X,i). As a result of the insertion of X,
(represented by \emph{(+)}), the code \emph{m1} is transformed to
m2. Then Transform\_triple calculates a new precondition for
\emph{m2} using the \emph{wp} calculus starting from the last line
of \emph{m2} and calculates a new postcondition using the
\emph{sp} calculus. The result is an intermediate triple \{P2\} m2
\{Q2\} that will be taken as an argument in the recursive call
with the remaining patch \emph{patch2}. The algorithm stops when
the patch is empty and raises exceptions when errors occur in the
calculus.

\item \textbf{Step (3):} \emph{\textbf{Implication proof}} The
calculated triple needs to be matched to the target triple to
establish the correctness of the transformation. The property that
needs to be shown here is implication. We show that the calculated
postcondition implies the target
 postcondition and that the target precondition implies the calculated postcondition: $ Q3\Rightarrow Q2 \bigwedge P2 \Rightarrow P3 $.
 \end{itemize}

 \section{Related work}
Several studies have been conducted in order to use static
semantics to prevent type errors in bytecode. Our work extends the
formalism presented in \cite{Freund:1999:TSO:330643.330646}. This
work defined semantics and a type system to study object
initialization in bytecode. The original idea was developed in
\cite{DBLP:journals/toplas/StataA99}  to study bytecode
subroutines. In \cite{DBLP:journals/jar/FreundM03}, the authors
extended the work \cite{Freund:1999:TSO:330643.330646} to bytecode
subroutines, virtual method invocation and exceptions. On the
behavioral side, using predicate transformation to reason about
bytecode properties has been studied in
\cite{Gregoire:2007:CVC:1793574.1793580} . The authors presented a
verification condition generator for bytecode formalized in the
Coq proof assistant and based on weakest precondition calculus.
Another work using wp to generate verification conditions from an
annotated bytecode is presented in \cite{DBLP:conf/sac/BurdyP06}
\cite{DBLP:conf/fase/BurdyHP07}. The use of strongest
postcondition calculus is not as popular as the wp calculus. A
study is presented in \cite{DBLP:journals/ase/GannodC96} as a
basis for formal reverse engineering for an imperative language.
Our work is close to \cite{Freund:1999:TSO:330643.330646} in the
sense of the use of static semantics to analyze bytecode. The
specificity of our work is the definition of semantics for
updates. We use predicate transformation to reason about bytecode
properties using existing tools for specification and proofs. Our
framework uses both weakest precondition and strongest
postcondition to reason about transformations.

\section{Conclusion and future work}

In this paper we propose a general framework for a formalization,
verification and reasoning about Java bytecode transformation. We
gave first an approach for verification by analyzing the modified
bytecode to ensure absence of type errors. We gave then an
approach for reasoning about bytecode transformation by using
predicate transformations.  The aim of the two methods combined is
to provide a complete framework that provides the two aspects:
static and behavioral. The second method focuses on behavioral
aspects and aims to the definition to a rich assertion language to
capture dynamic update features and effects on execution
structures such as frames and objects in the heap ( in a Java Card
virtual machine for example). These structures are not available
in the static aspect of the framework.

This work is on-going. Our aim immediately is to complete the
implementation by extending the language to other instructions in
bytecode and to the other possible transformations for methods
(adding arguments for example). On the other side, we aim to
complete the work concerning  behavioral aspects by defining
algorithms to take into account deleting instructions in predicate
transformation and to choose
 a configuration of existing tools for specification and reasoning. The
verification presented is implemented using the functional
language Ocaml. We aim to use mathematical reasoning to prove its
correctness. In the longer term, we wish to use a proof assistant
to reason about bytecode transformation.

\bibliographystyle{eptcs}
\bibliography{mabib}

\textbf\textsc{{APPENDIX: More rules for static semantics}}\\

\textbf{A. For instructions addition\\}

$ \frac{
\begin{array}{ll}
Add\_inst \ \  goto\ L (i+1)\\
SD_{i+1}=SD_{i}\\
S_{i+1}=S_{i} \ \ F_{i+1} =F_{i}\  PC\_MAX++\ \  \\
M2 = Add\_inst (M1, goto \ L, i+1)\\
 i+1, L \in DOM(BC)\
\end{array}
}{<(F_{i},S_{i},SD_{i},M1),i>\rightarrow\
<(F_{i+1},S_{i+1},SD_{i+1},M2),i+2>} \ \ \frac{
\begin{array}{ll}
Add\_inst \ \  pop \  (i+1)\\
SD_{i+1}=SD_{i}-1\\
S_{i}=t.S_{0}\rightarrow S_{i+1}=S_{0} \ \ \ F_{i+1} =F_{i}\\
M2 = Add\_inst (M1, pop, i+1)\\
PC\_MAX++\ \ \ \ i+1\in DOM(BC)
\end{array}
}{<(F_{i},S_{i},SD_{i},M1),i>\rightarrow
<(F_{i+1},S_{i+1},SD_{i+1},M2),i+2>} \ \ \
 $
\newline
 $ \frac{
\begin{array}{ll}
Add\_inst \ \  store \ x (i+1)\\
SD_{i+1}=SD_{i}-1 \ \ PC\_MAX++\\\
S_{i}= t.S_{0} \ \ F_{i+1} =F_{i}[x\leftarrow t] \ S_{i+1}=S_{0} \\
M2 = Add\_inst (M1, store \ x, i+1) \\
 i+1\in DOM(BC)\ x\in VAR(BC) \
\end{array}
}{<(F_{i},S_{i},SD_{i},M1),i>\rightarrow\
<(F_{i+1},S_{i+1},SD_{i+1},M2),i+2>} \ \ \frac{
\begin{array}{ll}
\ \\
Add\_inst\ putfield(A,f,t)  (i+1)\\
SD_{i+1}=SD_{i}-2\\
S_{i}=t.A.S_{0}\Rightarrow S_{i+1}=S_{0} \\
M2 = Add\_inst (M1, putfield (A,f,t), i+1)\\
PC\_MAX+3\ \ \ \ F_{i+1} =F_{i}\ \ i+1\in DOM(BC)
\end{array}
}{<(F_{i},S_{i},SD_{i},M1),i>\rightarrow
<(F_{i+1},S_{i+1},SD_{i+1},M2),i+3 >} \ \ \ \\$
 \newline
$ \frac{
\begin{array}{ll}
Add\_inst \ \  invokevirtuel(A,l,t)  (i+1)\\
SD_{i+1}=SD_{i} - (card^{*} (dom^{*} (t)) +1)\\
S_{i+1}=tn_{1}.tn_{2}\dots tn_{n} .S_{0} \rightarrow S_{i+1} =S_{0}\\
M2 = Add\_inst (M1, invokevirtuel(A,l,t), i+1)\\
 i+1\in DOM(BC)\ \ F_{i+1} =F_{i} \ \ PC\_MAX+3
\end{array}
}{<(F_{i},S_{i},SD_{i},M1),i>\rightarrow\
<(F_{i+1},S_{i+1},SD_{i+1},M2),i+2>} \ \ \ \\ $

\textbf{*Notations}
\begin{itemize}
\item \textbf{dom}: represents the domain of the invoked function
(types of its arguments)
\item \textbf{card:} represents the number of elements in the
domain.
\end{itemize}
\textbf{B. For instructions suppression}\\

$ \frac{
\begin{array}{ll}
Dlt\_inst \ \  pop \ \  (i+1))\\
SD_{i}= a \rightarrow SD_{i+1}=Effects\_SD^{**} (a, M2[i+1])\\
S_{i}=t.S_{0}\rightarrow [M2]S_{i+1}=Effects\_STK^{**} (M2[i+1],t.S_{0})\\
(M2)^{*}F_{i+1}= Effects\_F^{**} (M2[i+1],F_{i}) \\
M2 = Dlt\_inst (M1, pop , i+1)\\
i+1\in DOM(BC)\ PC\_MAX--\ \
\end{array}
}{<(F_{i},S_{i},SD_{i},M1),i>\rightarrow
<(F_{i+1},S_{i+1},SD_{i+1},M2),i+Instr\_length (M2 [1+i])>}\\ $

$ \frac{
\begin{array}{ll}
Dlt\_inst \ \  new\ A \ \  (i+1))\\
SD_{i}= a \rightarrow SD_{i+1}=Effects\_SD (a, M2[i+1])\\
S_{i}=t.S_{0}\rightarrow [M2]S_{i+1}=Effects\_STK (M2[i+1],t.S_{0})\\
(M2)F_{i+1}= Effects\_F (M2[i+1],F_{i}) \\
M2 = Dlt\_inst (M1, new \ A , i+1)\\
i+1\in DOM(BC)\ PC\_MAX--\ \
\end{array}
}{<(F_{i},S_{i},SD_{i},M1),i>\rightarrow
<(F_{i+1},S_{i+1},SD_{i+1},M2),i+Instr\_length (M2 [1+i])>}\\ $

$ \frac{
\begin{array}{ll}
Dlt\_inst \ \  load\ x \ \  (i+1))\\
SD_{i}= a \rightarrow SD_{i+1}=Effects\_SD (a, M2[i+1])\\
S_{i}=S_{0}\rightarrow (M1)S_{i+1}= t.S_{0},\ (M1)F_{i+1} (x) =t \rightarrow \\
(M2)S_{i+1}Effects\_STK (M2[i+1],S_{0})\\
(M2)F_{i+1}= Effects\_F (M2[i+1],F_{i}) \\
M2 = Dlt\_inst (M1, load \ x , i+1)\\
i+1\in DOM(BC)\ PC\_MAX--\ \
\end{array}
}{<(F_{i},S_{i},SD_{i},M1),i>\rightarrow
<(F_{i+1},S_{i+1},SD_{i+1},M2),i+Instr\_length (M2 [1+i])>}\\ $

\textbf{**Notations:}
\begin{itemize}
\item Effect\_STK ( a,b): represents the effect of the
instruction a on the stack a.
\item Effect\_F(a,b): represents the effect on the
instruction a on F.
\item Effect\_SD(a,b): represents the effect of then
instruction b on the stack depth a.
\item (M2)F: represents F according to the mapping M2.
\end{itemize}

\end{document}